\documentclass[12pt]{article}
\usepackage{epsf}
\usepackage{amsmath}
\usepackage{cite}
\usepackage{graphics}
\usepackage{amsfonts}
\usepackage{amssymb}
\usepackage{latexsym}
\usepackage{color}
\input{colordvi.tex}

\renewcommand{\thefootnote}{\#\arabic{footnote}}

\begin{document}
\setcounter{footnote}{0}

\begin{titlepage}

\begin{center}

\hfill HIP-2009-23/TH\\

\vskip .5in

{\Large \bf Effect of Background Evolution \\ on the Curvaton Non-Gaussianity
}

\vskip .45in

{\large
Kari Enqvist$\,^{1,2}$ and Tomo Takahashi$\,^3$
}

\vskip .45in

{\em
$^1$
Helsinki Institute of Physics, University of Helsinki, PO Box 64, FIN-00014,
Finland \\
$^2$
Department of Physical Science, University of Helsinki, \\PO Box 64, FIN-00014,
Finland \\
$^3$
Department of Physics, Saga University, Saga 840-8502, Japan
}

\end{center}

\vskip .4in

\begin{abstract}

  We investigate how the background evolution affects the curvature
  perturbations generated by the curvaton, assuming a curvaton
  potential that may deviate slightly from the quadratic one, and
  parameterizing the background fluid density as $\rho\propto
  a^{-\alpha}$, where $a$ is the scale factor, and $\alpha$ depends on
  the background fluid.  It turns out that the more there is deviation
  from the quadratic case, the more pronounced is the dependence of
  the curvature perturbation on $\alpha$. We also show that the
  background can have a significant effect on the nonlinearity
  parameters $f_{\rm NL}$ and $g_{\rm NL}$. As an example, if at the
  onset of the curvaton oscillation there is a dimension 6
  contribution to the potential at 5 \% level and the energy fraction
  of the curvaton to the total one at the time of its decay is at 1\%,
  we find variations $\Delta f_{\rm NL} \sim \mathcal{O}(10)$ and
  $\Delta g_{\rm NL} \sim \mathcal{O}(10^4)$ between matter and
  radiation dominated backgrounds.  Moreover, we demonstrate that
  there is a relation between $f_{\rm NL}$ and $g_{\rm NL}$ that can
  be used to probe the form of the curvaton potential and the equation
  of state of the background fluid.

\end{abstract}
\end{titlepage}

\renewcommand{\thepage}{\arabic{page}}
\setcounter{page}{1}
\renewcommand{\thefootnote}{\#\arabic{footnote}}

\section{Introduction}

Although quantum fluctuations of the inflaton are often taken to be
responsible for the origin of density perturbations, other mechanisms
such as the curvaton \cite{Enqvist:2001zp,Lyth:2001nq,Moroi:2001ct},
where fluctuations of a scalar field other than the inflaton generate
primordial perturbations, have also attracted much attention recently.
In particular, in the light of the recent result from WMAP5 which
suggests that primordial non-Gaussianity may be large
\cite{Komatsu:2008hk,Smith:2009jr}\footnote{
  The degree of non-Gaussianity is usually characterized by the lowest
  order non-linearity parameter $f_{\rm NL}$. The current constraint
  on $f_{\rm NL}$ is $-9 < f_{\rm NL} < 111$ in
  Ref.~\cite{Komatsu:2008hk} or $-4 < f_{\rm NL} < 80$ at $95\%$
  C.L. in Ref.~\cite{Smith:2009jr}.  Although purely Gaussian
  fluctuations with $f_{\rm NL}=0$ are allowed, the central value is
  away from zero.
}, the curvaton mechanism may be attractive since a large primordial
non-Gaussianity can be generated in this scenario
\cite{Lyth:2002my,Bartolo:2003jx,Enqvist:2005pg,Malik:2006pm,Sasaki:2006kq,Huang:2008ze,Ichikawa:2008iq,Multamaki:2008yv,Li:2008jn,Enqvist:2008gk,Huang:2008bg,Huang:2008zj,Moroi:2008nn,Kawasaki:2008mc,Chingangbam:2009xi}\footnote{
  Other models, for example, such as the modulated reheating scenario
  \cite{Dvali:2003em,Kofman:2003nx} are also known to generate large
  non-Gaussian fluctuations
  \cite{Zaldarriaga:2003my,Suyama:2007bg,Ichikawa:2008ne,Takahashi:2009cx}.
}, whereas simplest inflation models predict a non-linearity parameter
$f_{\rm NL}$ that is of the order of the slow-roll parameters (or
{\cal O}(1) at most), and hence practically imply a Gaussian
perturbation.

Usually the curvaton potential is assumed to be quadratic.  However,
there is no other reason for this except simplicity, and in fact in
any realistic particle physics model the curvaton can be expected to
have some self-interactions. Thus, deviations from the exact quadratic
form are also worth investigating.  They have been discussed in
\cite{Enqvist:2005pg,Enqvist:2008gk,Huang:2008bg,Huang:2008zj,Kawasaki:2008mc,Chingangbam:2009xi,dynamics,Enqvist:2009zf},
where it has been pointed out that non-quadratic contributions to the
potential can modify the resultant curvaton perturbations in a
significant manner. In particular, the prediction for the
non-linearity parameters $f_{\rm NL}$ and $g_{\rm NL}$ can change
considerably as compared to the quadratic case.

In addition, there is yet another assumption which is tacitly adopted
in the curvaton literature: the background evolution of the universe
is determined by radiation. With this assumption, the curvaton starts
to oscillate during a radiation-dominated (RD) epoch. If the curvaton
decays before dominating the Universe, radiation is always the
dominant component and controls the background evolution of the
Universe.  However, it is also possible that after inflation, the
inflaton is oscillating around the minimum of the potential for a
while and that the curvaton begins to oscillate during such epoch. In
this case, the background evolution is different from the case of
radiation and is determined by a matter-like component if the inflaton
potential is approximatively quadratic\footnote{
  In Ref.~\cite{Moroi:2008nn}, this kind of situation is also included
  in the analysis for the quadratic curvaton potential.
}.  After inflation there could also exist a possibility of a
kination-dominated phase where the kinetic term of some scalar field
can dominate the energy density of the universe.  Such fluid has a
stiff equation of state with $w=1$ while its energy density decreases
as $\rho \propto a^{-6}$.

In this paper, we investigate how the background evolution of the
universe affects the curvature perturbation generated from the
curvaton.  We first show that when the curvaton potential has a
quadratic form, the background evolution has little effect on the
curvature perturbation.  However, when the curvaton potential includes
a non-quadratic term, the nonlinear evolution of the curvaton field is
affected much by the background, and the resultant curvature
perturbations can be significantly modified from the usual RD case.
We investigate this issue by assuming a general background fluid and a
potential that slightly deviates from a quadratic form, and derive the
dependence of the curvature perturbation and/or the non-linearity
parameters such as $f_{\rm NL}$ and $g_{\rm NL}$ on different
background fluids.

The structure of the paper is as follows. In the next section, we
summarize the formalism and give definitions of the quantities
required for the subsequent analysis. In section~\ref{sec:effect}, we
discuss how the background evolution affects the quantities such as
power spectrum and non-linearity parameters, using the appropriate
formulas for arbitrary background fluids.  The final section is
devoted to a summary and a discussion of the results.

\section{Formalism and Definitions}

Let us begin by summarizing the formalism and the definitions of the
various quantities required for the subsequent discussion.  Here we
consider a potential of the curvaton field $\sigma$ which, in addition
to the usual quadratic term, also includes a non-renormalizable term:
\begin{equation}
\label{eq:V}
V(\sigma)
=
\frac{1}{2} m_\sigma^2 \sigma^2
+
\lambda m_\sigma^4 \left( \frac{\sigma}{m_\sigma} \right)^n~,
\end{equation}
where $m_\sigma$ is the mass of the curvaton, and $\lambda$ is a
constant. For the purpose of this paper, it is enough to investigate
the case of a slight deviation from the purely quadratic form. Thus in
the following we assume that the quadratic term always dominates over
the non-quadratic term (for a general discussion of the ramifications
of non-quadratic terms in curvaton models, see \cite{dynamics,
  Enqvist:2009zf}).  We characterize the relative contribution of the
non-quadratic term at the time when the curvaton is still in a
slowly-rolling regime by the parameter $s$, defined as
\begin{equation}
\label{eq:def_s}
s \equiv 2 \lambda \left( \frac{\sigma_\ast}{m_\sigma} \right)^{n-2}.
\end{equation}

To investigate the curvature perturbation $\zeta$ generated by the
curvaton field, we adopt the $\delta N$ formalism
\cite{Starobinsky:1986fxa,Sasaki:1995aw,Sasaki:1998ug,Lyth:2004gb} and
calculate $\zeta$ up to the third order as
\begin{equation}
\label{eq:defs}
\zeta =
\frac{dN}{d\sigma_\ast} \delta \sigma_\ast
+ \frac{1}{2} \frac{d^2N}{d\sigma_\ast^2} (\delta \sigma_\ast)^2
+ \frac{1}{6} \frac{d^3N}{d\sigma_\ast^3} (\delta \sigma_\ast)^3
+ \cdots.
\end{equation}
Once we obtain $\zeta$ up to the third order, the power spectrum
$P_\zeta$, bispectum $B_\zeta $, and trispectrum $T_\zeta $ are given
by
\begin{equation}
\label{eq:power}
\langle \zeta_{\vec k_1} \zeta_{\vec k_2} \rangle
=
{(2\pi)}^3 P_\zeta (k_1) \delta ({\vec k_1}+{\vec k_2}),
\end{equation}
\begin{eqnarray}
\langle \zeta_{\vec k_1} \zeta_{\vec k_2} \zeta_{\vec k_3} \rangle
&=&
{(2\pi)}^3 B_\zeta (k_1,k_2,k_3) \delta ({\vec k_1}+{\vec k_2}+{\vec k_3}).
\label{eq:bi}
\end{eqnarray}
\begin{eqnarray}
\langle
\zeta_{\vec k_1} \zeta_{\vec k_2} \zeta_{\vec k_3} \zeta_{\vec k_4}
\rangle
&=&
{(2\pi)}^3 T_\zeta (k_1,k_2,k_3,k_4) \delta ({\vec k_1}+{\vec k_2}+{\vec k_3}+{\vec k_4}),
\label{eq:tri}
\end{eqnarray}
where $B_\zeta$ and $T_\zeta$ can be written as
\begin{eqnarray}
B_\zeta (k_1,k_2,k_3)
&=&
\frac{6}{5} f_{\rm NL}
\left(
P_\zeta (k_1) P_\zeta (k_2)
+ P_\zeta (k_2) P_\zeta (k_3)
+ P_\zeta (k_3) P_\zeta (k_1)
\right), \\
\label{eq:def_f_NL}
T_\zeta (k_1,k_2,k_3,k_4)
&=&
\tau_{\rm NL} \left(
P_\zeta(k_{13}) P_\zeta (k_3) P_\zeta (k_4)+11~{\rm perms.}
\right) \nonumber \\
&&
+ \frac{54}{25} g_{\rm NL} \left( P_\zeta (k_2) P_\zeta (k_3) P_\zeta (k_4)
+3~{\rm perms.} \right).
\label{eq:def_tau_g_NL}
\end{eqnarray}
Here $f_{\rm NL}, \tau_{\rm NL}$ and $g_{\rm NL}$ are non-linearity
parameters often used in the literature.  Note that, in our case,
$\tau_{\rm NL}$ is related to $f_{\rm NL}$ by
\begin{equation}
\label{eq:taurelated}
\tau_{\rm NL} = \frac{36}{25} f_{\rm NL}^2~.
\end{equation}
Thus, alternatively, we may assume the following expansion as the
definition of $f_{\rm NL}$ and $g_{\rm NL}$
\begin{equation}
\label{eq:defs}
\zeta = \zeta_1 + \frac{3}{5} f_{\rm NL} \zeta_1^2 + \frac{9}{25} g_{\rm NL}\zeta_1^3
+ \cdots,
\end{equation}
where $\zeta_1$ denotes the curvature perturbation at linear order.

Let us now derive the formulae for the power spectrum, $f_{\rm NL}$
and $g_{\rm NL}$ for a general background equation of state. As we
mentioned in the Introduction, usually one assumes that the curvaton
oscillates in a radiation background evolution.  However, it is
obvious that the inflaton can be expected to oscillate around the
minimum of its (quadratic) potential for some time before it decays.
For a weakly coupled inflaton field, the duration of this epoch could
be fairly long.  During that time the inflaton behaves like matter,
and it is possible that the curvaton oscillations begin during this
epoch.

Eventually the inflaton will decay into radiation; this can well
happen before the curvaton oscillations dominate the energy budget of
the universe, or before the curvaton decay, whichever happens first.
Hence the universe becomes radiation-dominated so that the background
evolution for oscillating curvaton is controlled first by a matter
component, followed by radiation domination. Having this kind of
situation in mind, let us assume there is a transition in the
background from one fluid to another one at time $t_{\rm tr}$.  We
then parametrize the energy density of the background fluid before the
transition time by
\begin{equation}
\rho_{\rm BG} \propto a^{-\alpha},
\end{equation}
whereas after the transition, for $ t > t_{\rm tr}$, we have
\begin{equation}
\rho_{\rm BG} \propto a^{-\beta}.
\end{equation}
For the case of oscillating inflaton background, the transition epoch
$t_{\rm tr}$ corresponds to the time when the inflaton decays.  In
this case, for $t < t_{\rm tr}$, $\alpha = 3$ while for $t > t_{\rm
  tr}$, $\beta = 4$.  Another example is a universe that is first
kination-dominated so that the kinetic energy of a scalar field
dominates the energy density of the universe. Then, after some time,
the universe becomes radiation-dominated. In this case, $\alpha=6$ for
$t < t_{\rm tr}$ while after the transition time $\beta=4$.

With this parametrization, we obtain the curvature perturbation at the
linear order as
\begin{eqnarray}
\zeta_1=  \frac{2 \sigma'_{\rm ocs} }{3 \sigma_{\rm osc}} R
\delta \sigma~,
\end{eqnarray}
where
\begin{equation}
R \equiv r_{\rm dec} \left( 1- k r_{\rm tr} \right)  + k r_{\rm tr}.
\end{equation}
The prime denotes the derivative with respect to $\sigma_\ast$.  Here
$r_{\rm dec}$ and $r_{\rm tr}$ roughly correspond to the fraction of
energy density of the curvaton to the total energy density at the time
of the curvaton decay and the transition of the background,
respectively. Their precise definitions are
\begin{equation}
\label{eq:def_r}
\left.
r_{\rm tr}
\equiv \frac{3 \rho_\sigma}{\alpha \rho_{\rm BG} + 3\rho_\sigma}\right|_{\rm tr},~~~~
\left.
r_{\rm dec}
\equiv \frac{3 \rho_\sigma}{\beta \rho_{\rm BG} + 3\rho_\sigma}\right|_{\rm dec}.
\end{equation}
Furthermore, $k$ is defined by
\begin{equation}
k = 1- \frac{\alpha}{\beta}.
\end{equation}
By calculating the curvature perturbation up to the third order, we
find that the non-linearity parameters $f_{\rm NL}$ and $g_{\rm NL}$
read as
\begin{equation}
\label{eq:fNL_general}
f_{\rm NL} =
\frac{5}{6} \left[
\frac{3}{2R} \left(
\frac{\sigma^{\prime\prime}_{\rm osc} \sigma_{\rm osc}}{\sigma^{\prime 2}_{\rm osc}} -1
\right)
+ \frac{3}{2} \frac{\sigma_{\rm osc} }{R^2} \frac{dR}{d\sigma_{\rm osc}}
\right],
\end{equation}

\begin{equation}
\label{eq:fNL_general}
g_{\rm NL} =
\frac{25}{54} \left[
\frac{9}{4R^2} \left(
\frac{\sigma^{\prime\prime\prime}_{\rm osc} \sigma_{\rm osc}^2}
{(\sigma^{\prime})^3_{\rm osc}}
-
3\frac{\sigma^{\prime\prime}_{\rm osc} \sigma_{\rm osc}}
{(\sigma^{\prime}_{\rm osc})^2}
+2
\right)
+ \frac{9 \sigma_{\rm osc}}{2 R^3}  \frac{dR}{d\sigma_{\rm osc}}\left(
\frac{3}{2} \frac{\sigma^{\prime\prime}_{\rm osc} \sigma_{\rm osc}}
{(\sigma^{\prime}_{\rm osc})^2}
- 1
\right)
+ \frac{9 \sigma_{\rm osc}^2}{4R^3}\frac{d^2R}{d\sigma^2_{\rm osc}}
\right],
\end{equation}
Here the derivatives of $R$ with respect to $\sigma_{\rm osc}$ appear.
They are given in terms of the derivatives of $r_{\rm dec}$ and
$r_{\rm tr}$ with respect to $\sigma_{\rm osc}$ as
\begin{equation}
\label{eq:R_prime}
\frac{dR}{d\sigma_{\rm osc}}
=
( 1- k r_{\rm tr} ) \frac{d r_{\rm dec}}{d \sigma_{\rm osc}}
+
k( 1 - r_{\rm dec} ) \frac{d r_{\rm tr}}{d \sigma_{\rm osc}},
\end{equation}
and
\begin{equation}
\label{eq:R_2prime}
\frac{d^2R}{d\sigma_{\rm osc}^2}
=
( 1- k r_{\rm tr} ) \frac{d^2 r_{\rm dec}}{d \sigma_{\rm osc}^2}
- 2 k \frac{d r_{\rm dec}}{d \sigma_{\rm osc}} \frac{d r_{\rm dec}}{d \sigma_{\rm osc}}
+
k( 1 - r_{\rm dec} ) \frac{d^2 r_{\rm tr}}{d \sigma_{\rm osc}^2}.
\end{equation}
The derivatives $d r_{\rm dec} / d \sigma_{\rm osc} $ and $dr_{\rm tr}
/ d \sigma_{\rm osc}$ are explicitly given by
\begin{eqnarray}
\frac{d r_{\rm dec}}{d \sigma_{\rm osc}}
&= &
\frac{2}{3\sigma_{\rm osc}} r_{\rm dec} (1-r_{\rm dec} )
[ 3 + (\beta-3) r_{\rm dec} ] ( 1 -k r_{\rm tr} ), \\
\frac{d r_{\rm tr}}{d \sigma_{\rm osc}}
&=&
\frac{2}{3\sigma_{\rm osc}} r_{\rm tr} (1-r_{\rm tr} )
[ 3 + (\alpha-3) r_{\rm tr} ]
\end{eqnarray}
The second derivatives of $r_{\rm dec}$ and $r_{\rm tr}$ with respect
to $\sigma_{\rm osc}$ can be derived by differentiating the above
equations and they are calculated as
\begin{eqnarray}
\frac{d^2 r_{\rm dec}}{d \sigma_{\rm osc}^2}
&= &
\frac{2}{3\sigma_{\rm osc}} \frac{d r_{\rm dec}}{d \sigma_{\rm osc}}
\left[ -\frac{3}{2}
+ (1-2 r_{\rm dec} )
\{ 3 + (\beta-3) r_{\rm dec} \} ( 1 -k r_{\rm tr} )
\right. \notag \\
&&
\left.
+ r_{\rm dec}(1-r_{\rm dec}) (\beta - 3)( 1 -k r_{\rm tr} )
\right]
- \frac{2k}{3\sigma_{\rm osc}} \frac{d r_{\rm tr}}{d \sigma_{\rm osc}}
r_{\rm dec} (1-r_{\rm dec} )
\{ 3 + (\beta-3) r_{\rm dec} \}, \notag \\  \\ \notag \\
\frac{d^2 r_{\rm tr}}{d \sigma_{\rm osc}^2}
&=&
\frac{2}{3\sigma_{\rm osc}} \frac{d r_{\rm tr}}{d \sigma_{\rm osc}}
\left[ -\frac{3}{2}
+ (1-2 r_{\rm tr} )
\{ 3 + (\alpha-3) r_{\rm tr} \}
+(\alpha-3) r_{\rm tr} ( 1 - r_{\rm tr} )
\right].
\end{eqnarray}
We can now write down $f_{\rm NL}$ and $g_{\rm NL}$ as functions of
the parameters $\alpha, \beta, r_{\rm dec}$ and $r_{\rm tr}$ in an
explicit way, although in general the expressions are very
complicated.  In fact, as far as we consider only the cases with
$\alpha, \beta > 3$, $r_{\rm tr}$ should always be smaller than
$r_{\rm dec}$ and in most cases we may assume that $r_{\rm tr} \ll
r_{\rm dec}$. In this case, $R \simeq r_{\rm dec}$ and $\zeta_1,
f_{\rm NL}$ and $g_{\rm NL}$ may approximately be written as
 \begin{eqnarray}
\label{eq:zeta_apprx}
\zeta_1 &=&
\frac{2}{3} r_{\rm dec} \frac{\sigma'_{\rm osc}}{\sigma_{\rm osc}}  \delta \sigma_\ast,
\\
\label{eq:fNL_apprx}
f_{\rm NL} &=& \frac{5}{4r_{\rm dec}} \left(
1 +
\frac{\sigma_{\rm osc} \sigma_{\rm osc}^{\prime\prime}}{\sigma_{\rm osc}^{\prime 2}}
\right)
+ \frac{5}{6}(\beta-6) -\frac{5r_{\rm dec}}{6}(\beta-3), \\
\label{eq:gNL_apprx}
g_{\rm NL} &=& \frac{25}{54}
\left[
\frac{9}{4r_{\rm dec}^2}  \left(
\frac{\sigma_{\rm osc}^2 \sigma_{\rm osc}^{\prime\prime\prime}}
{\sigma_{\rm osc}^{\prime 3}}
+
3\frac{\sigma_{\rm osc} \sigma_{\rm osc}^{\prime\prime}}{\sigma_{\rm osc}^{\prime 2}}
\right)
+\frac{9}{2r_{\rm dec}}(\beta-6)
\left(
1
+
\frac{\sigma_{\rm osc} \sigma_{\rm osc}^{\prime\prime}}{\sigma_{\rm osc}^{\prime 2}}
\right)
\right. \notag \\
&&
\left.
+\frac{1}{2} \left(
189 - 63 \beta + 4\beta^2
-
9(\beta-3)
\frac{\sigma_{\rm osc} \sigma_{\rm osc}^{\prime\prime}}{\sigma_{\rm osc}^{\prime 2}}
\right)
-5r_{\rm dec} ( 18 -9 \beta +\beta^2)
 + 3r_{\rm dec}^2(\beta-3)^2
\right]. \notag \\
\end{eqnarray}
As one can notice from these expressions, when $r_{\rm dec}$ is small,
which is the interesting case because then non-Gaussianity can be
large, the effects of the background are almost encoded in the changes
of $\sigma_{\rm osc}$ and its derivatives. The quantities such as
$\sigma_{\rm osc}, \sigma_{\rm osc}^\prime$ and $\sigma_{\rm
  osc}^{\prime\prime}$ are evaluated at the beginning of the curvaton
oscillation, and thus the background evolution after the transition is
irrelevant if we assume a nearly quadratic potential for the curvaton.
Thus in this case, after its change the background evolution does not
affect much the curvature perturbation but modifies $\zeta_1$ and its
non-linearity parameters at most of order $\mathcal{O}(1)$. This can
be seen directly from the above expressions: $\beta$ affects the
coefficients by $\mathcal{O}(1)$.

If there is no change in the background evolution up to or until after
the time at which the curvaton decays, the expressions for the
curvature perturbation and non-linearity parameters are quite similar
to the case with $r_{\rm tr} \ll r_{\rm dec}$ mentioned above.  In
this case, we can simply set $k=0$ and $R=r_{\rm dec}$ in the above
equations. Then $\zeta_1, f_{\rm NL}$ and $g_{\rm NL}$ can be found by
replacing $\beta$ with $\alpha$ in Eqs.~\eqref{eq:zeta_apprx},
\eqref{eq:fNL_apprx} and \eqref{eq:gNL_apprx}.

\section{Background evolution and non-Gaussianity}\label{sec:effect}

Let us now apply the formalism of the previous Section and discuss how
the background evolution affects the amplitude of the curvature
perturbation and its non-linearity parameters. As we pointed out, the
transition in the background does not affect the perturbation much if
the transition occurs much before curvaton decay. Hence here we focus
on the case of no transition in the background, i.e., a single fluid
controls the background evolution from the time well before the
curvaton begins to oscillate to well after the curvaton has decayed.

\begin{figure}[htbp]
  \begin{center}
    \resizebox{170mm}{!}{
        \hspace{-20mm}
    \includegraphics{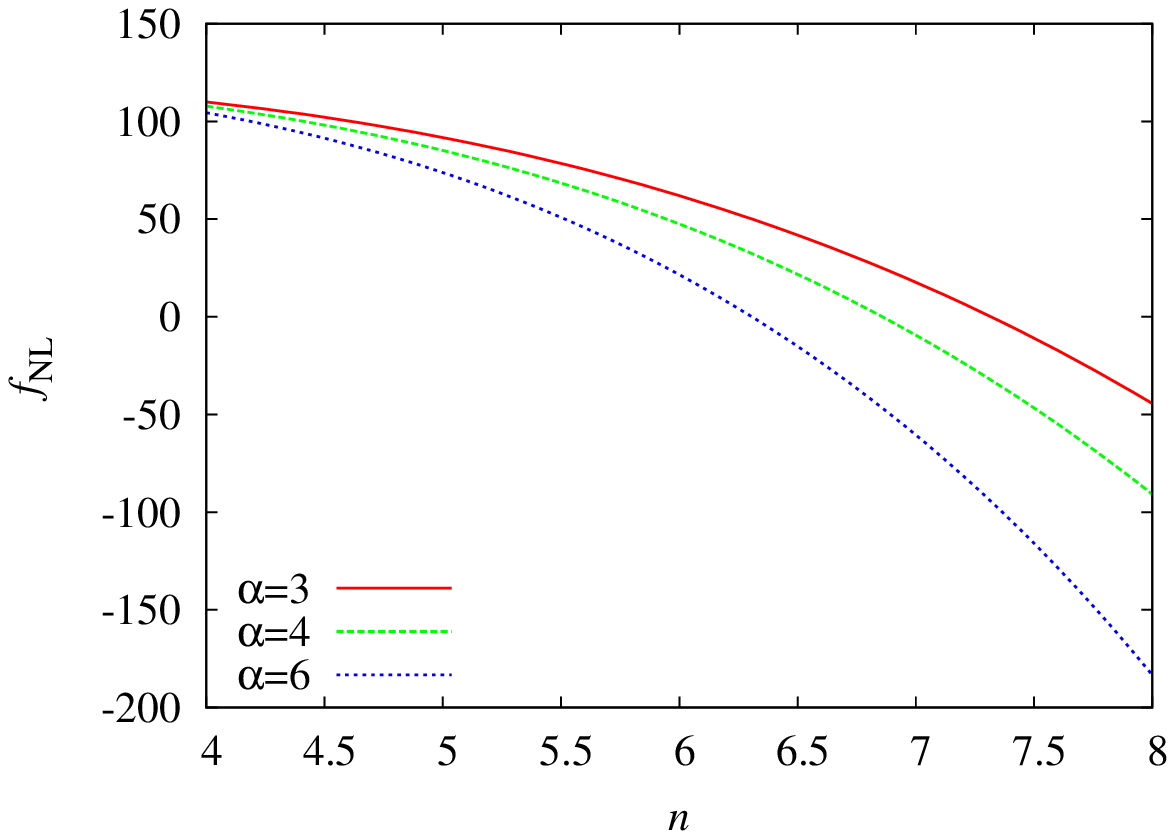}
    \includegraphics{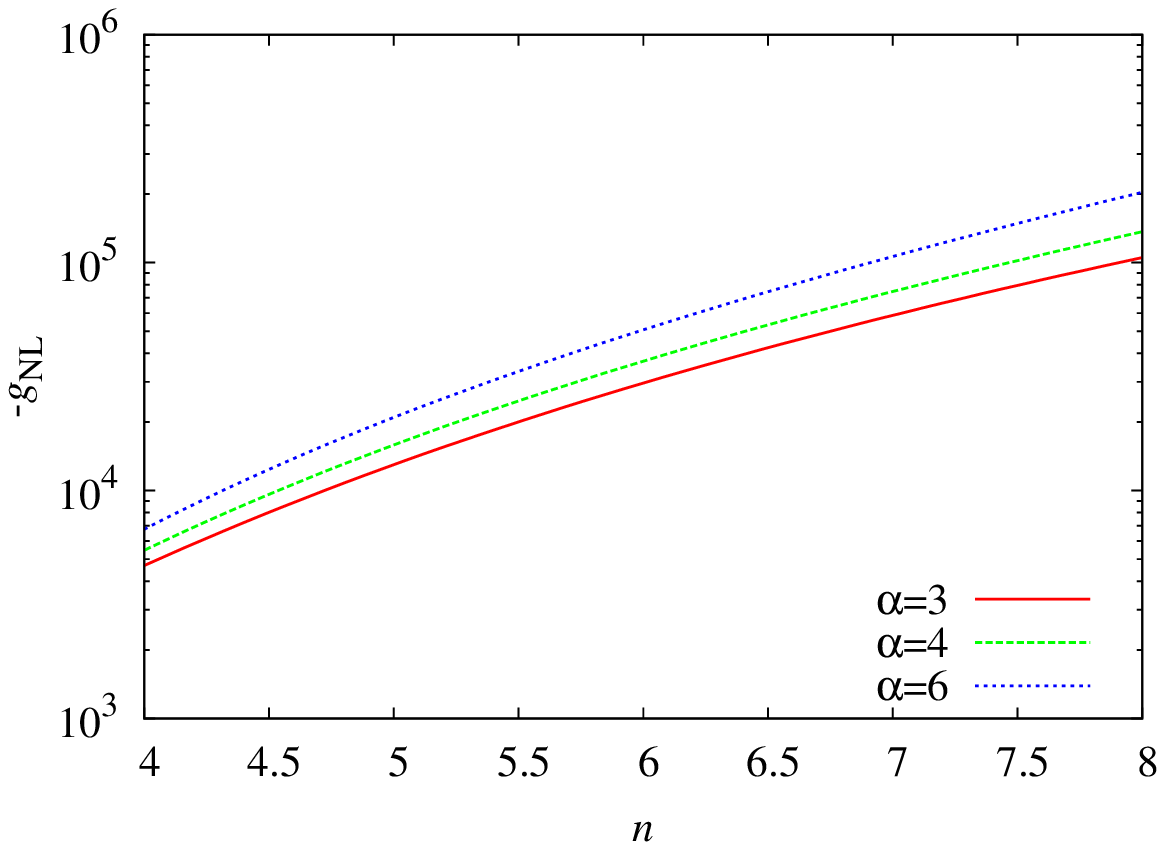}
    }
  \end{center}
  \caption{Plots of $f_{\rm NL}$ and $g_{\rm NL}$ as a function of $n$
    for several values of $\alpha$.  The values of $s$ and $r_{\rm
      dec}$ are taken as $s=0.05$ and $r_{\rm dec}=0.01$.}
  \label{fig:fNL_gNL_n}
\end{figure}

\begin{figure}[htbp]
  \begin{center}
    \resizebox{170mm}{!}{
        \hspace{-20mm}
    \includegraphics{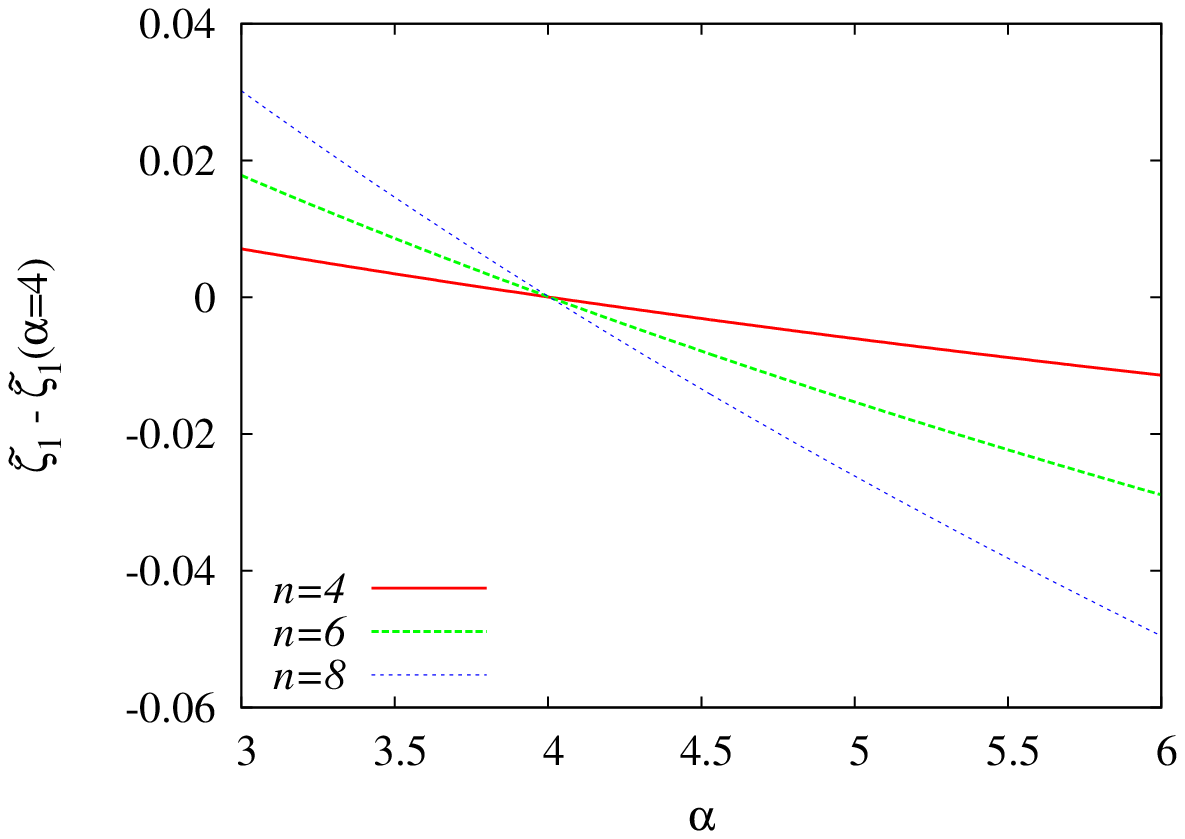}
    \includegraphics{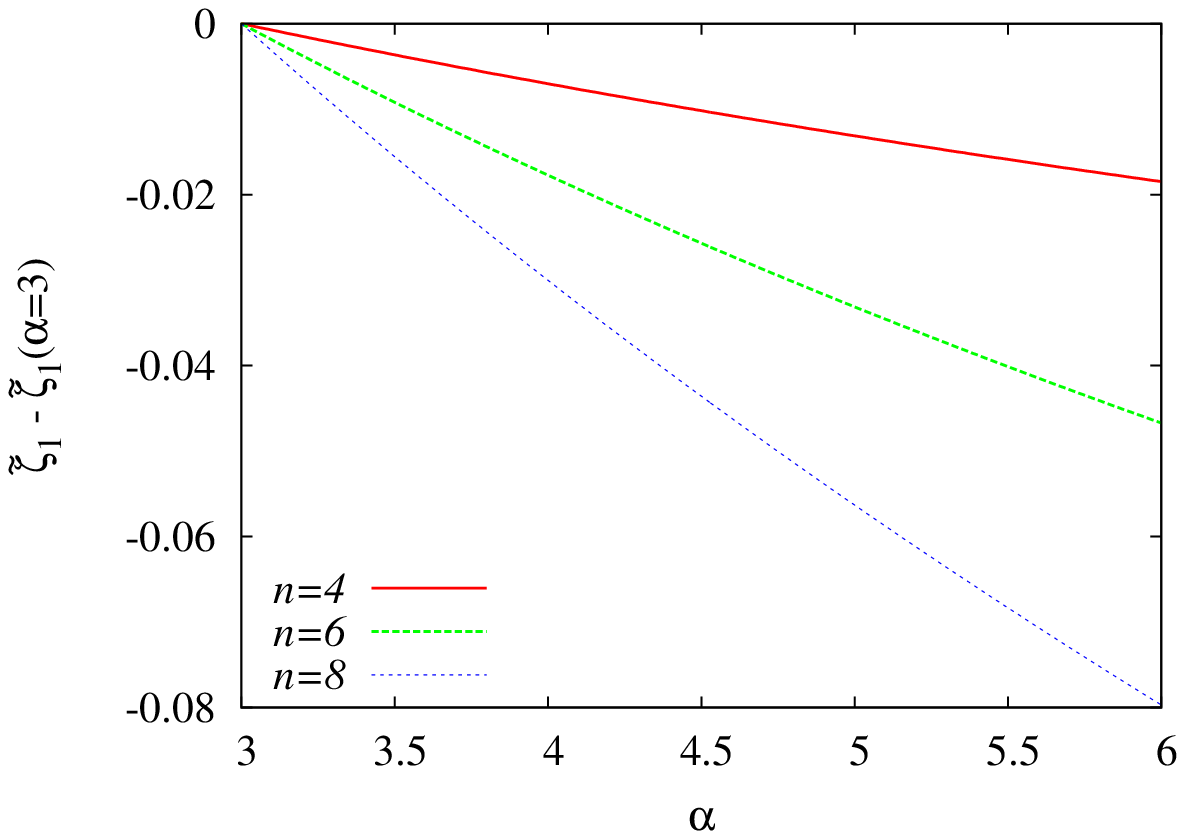}
    }
  \end{center}
  \caption{Plots of $\tilde{\zeta}_1 = \zeta_1 /\zeta_1^{\rm
      (quadratic)}$, which is normalized to $\zeta_1$ for the pure
    quadratic case, relative to that for the cases with $\alpha =4$
    (left) and $\alpha =3$ (right).  The values of $s$ and $r_{\rm
      dec}$ are taken as $s=0.05$ and $r_{\rm dec}=0.01$.}
  \label{fig:zeta_1}
\end{figure}

\begin{figure}[htbp]
  \begin{center}
    \resizebox{170mm}{!}{
        \hspace{-20mm}
    \includegraphics{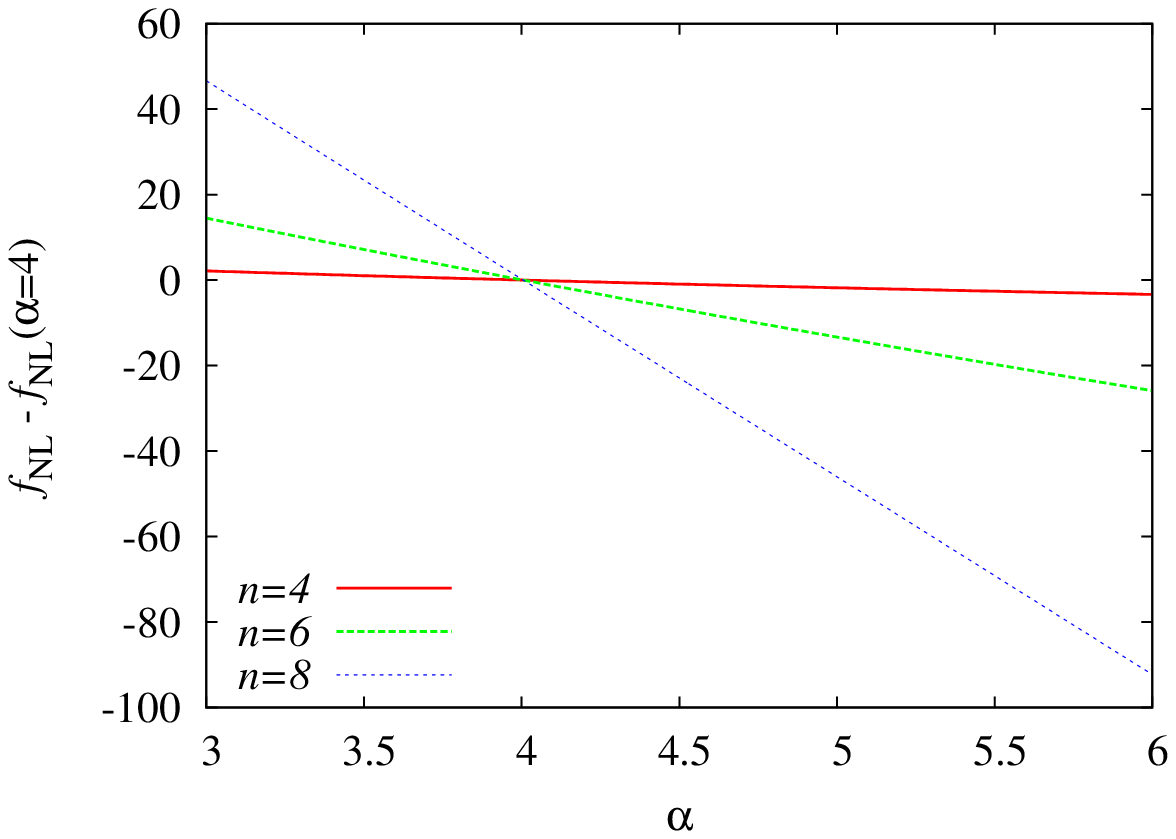}
    \includegraphics{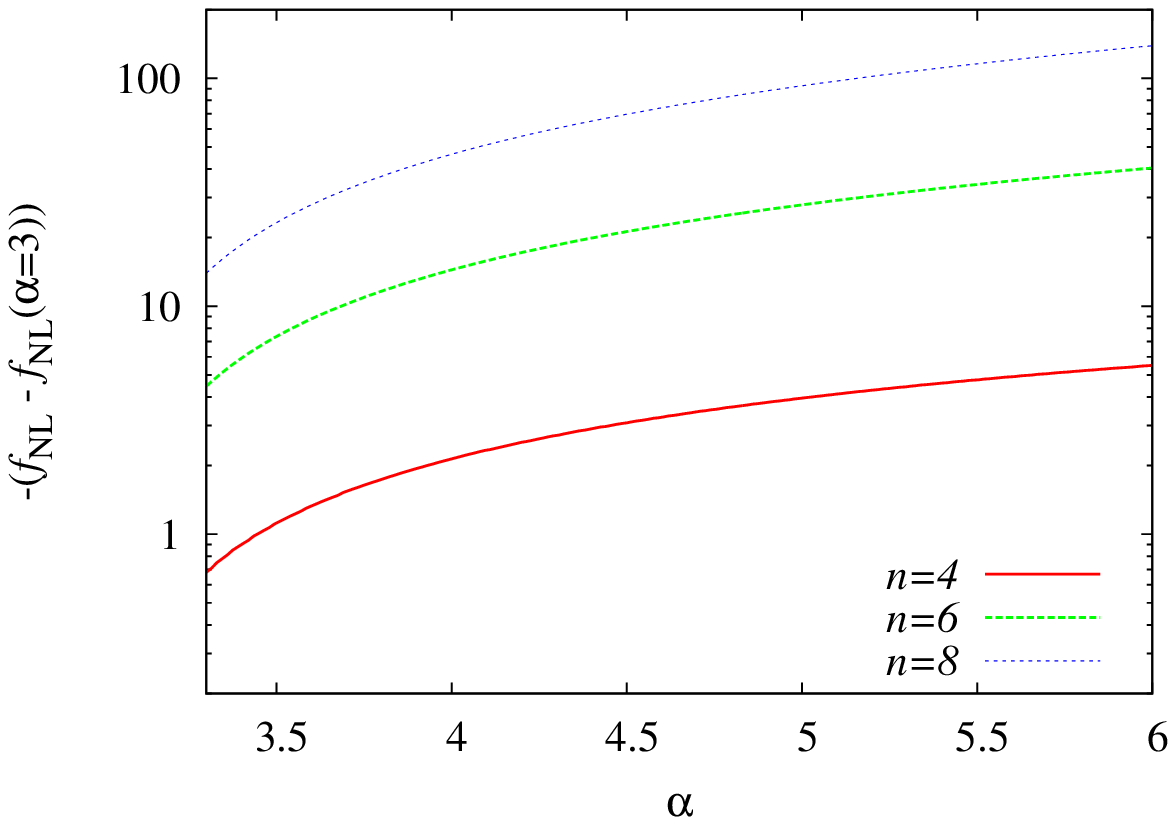}
    }
  \end{center}
  \caption{Plots of $f_{\rm NL}$ relative to that for the cases with
    $\alpha =4$ (left) and $\alpha =3$ (right).  The values of $s$ and
    $r_{\rm dec}$ are taken as $s=0.05$ and $r_{\rm dec}=0.01$.}
  \label{fig:fNL_1}
\end{figure}

\begin{figure}[htbp]
  \begin{center}
    \resizebox{170mm}{!}{
        \hspace{-20mm}
    \includegraphics{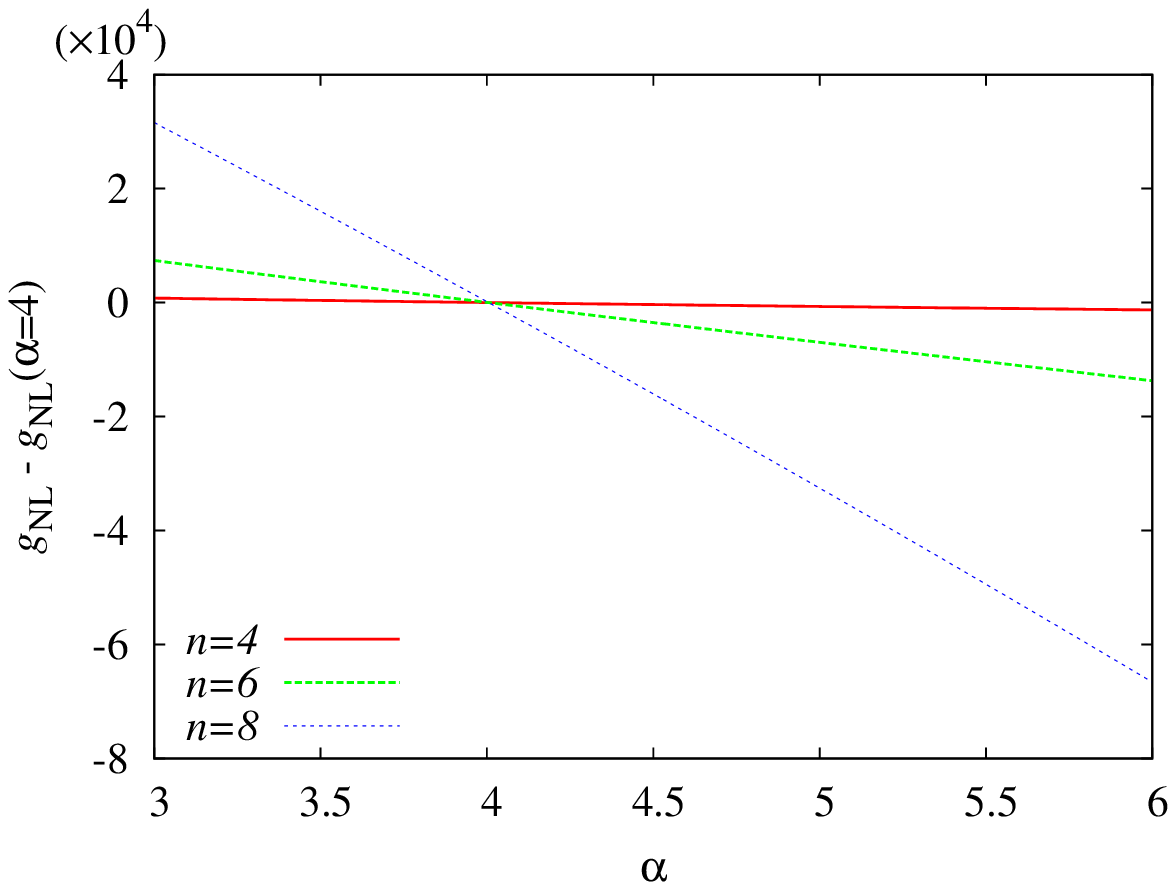}
    \includegraphics{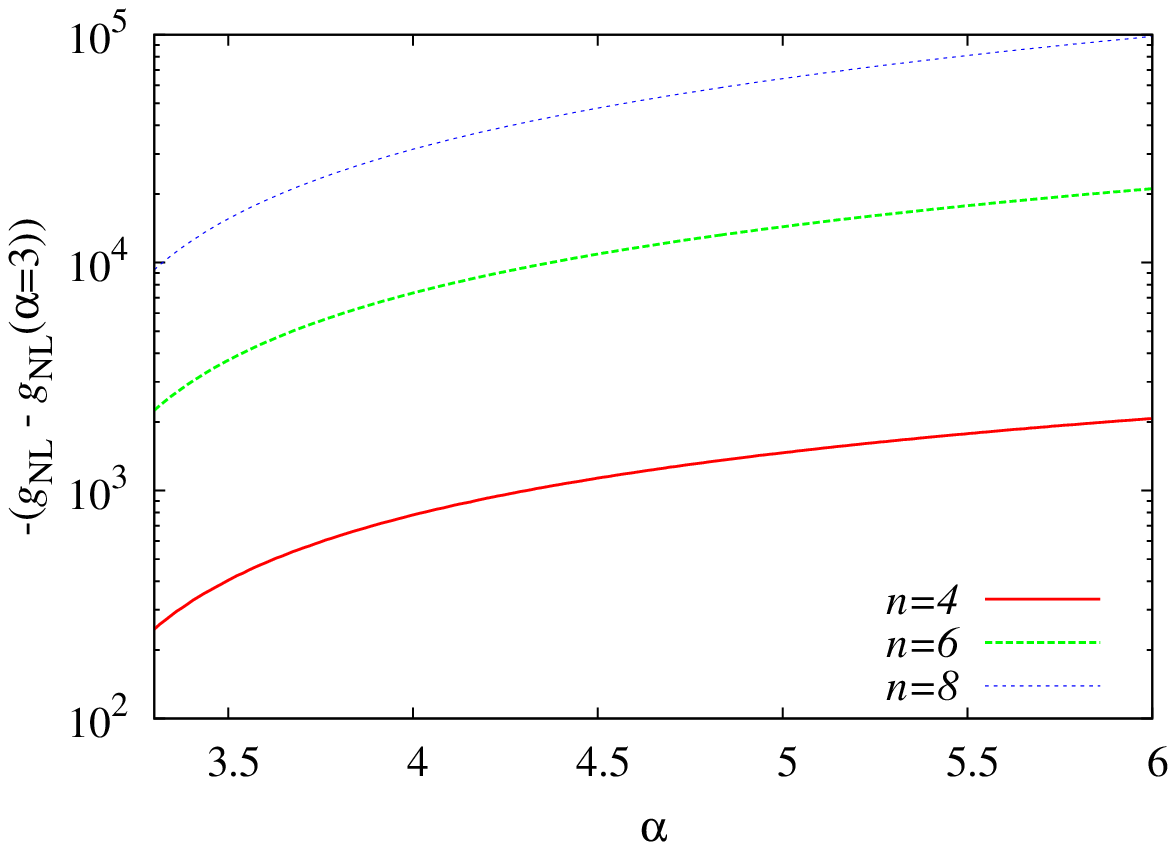}
    }
  \end{center}
  \caption{Plots of $g_{\rm NL}$ relative to that for the cases with
    $\alpha =4$ (left) and $\alpha =3$ (right).  The values of $s$ and
    $r_{\rm dec}$ are taken as $s=0.05$ and $r_{\rm dec}=0.01$.}
  \label{fig:gNL_1}
\end{figure}

\begin{figure}[htbp]
  \begin{center}
    \resizebox{170mm}{!}{
        \hspace{-20mm}
    \includegraphics{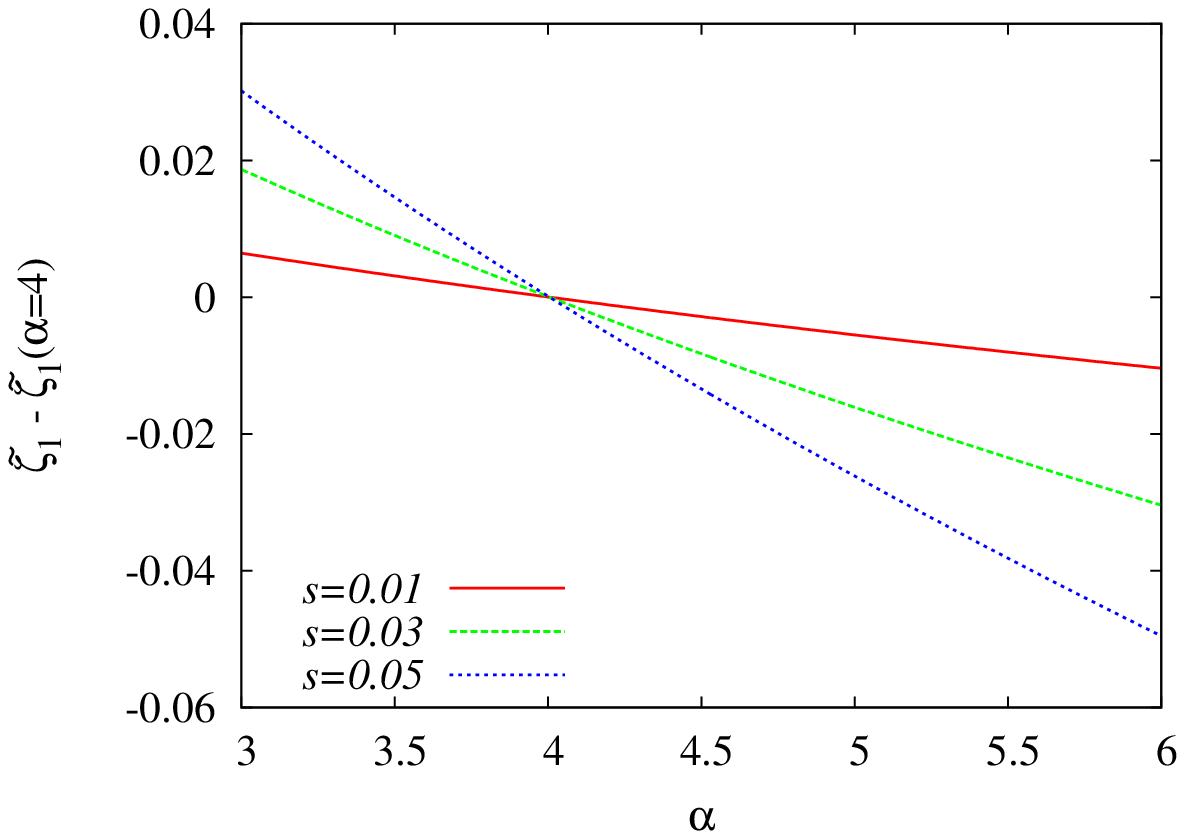}
    \includegraphics{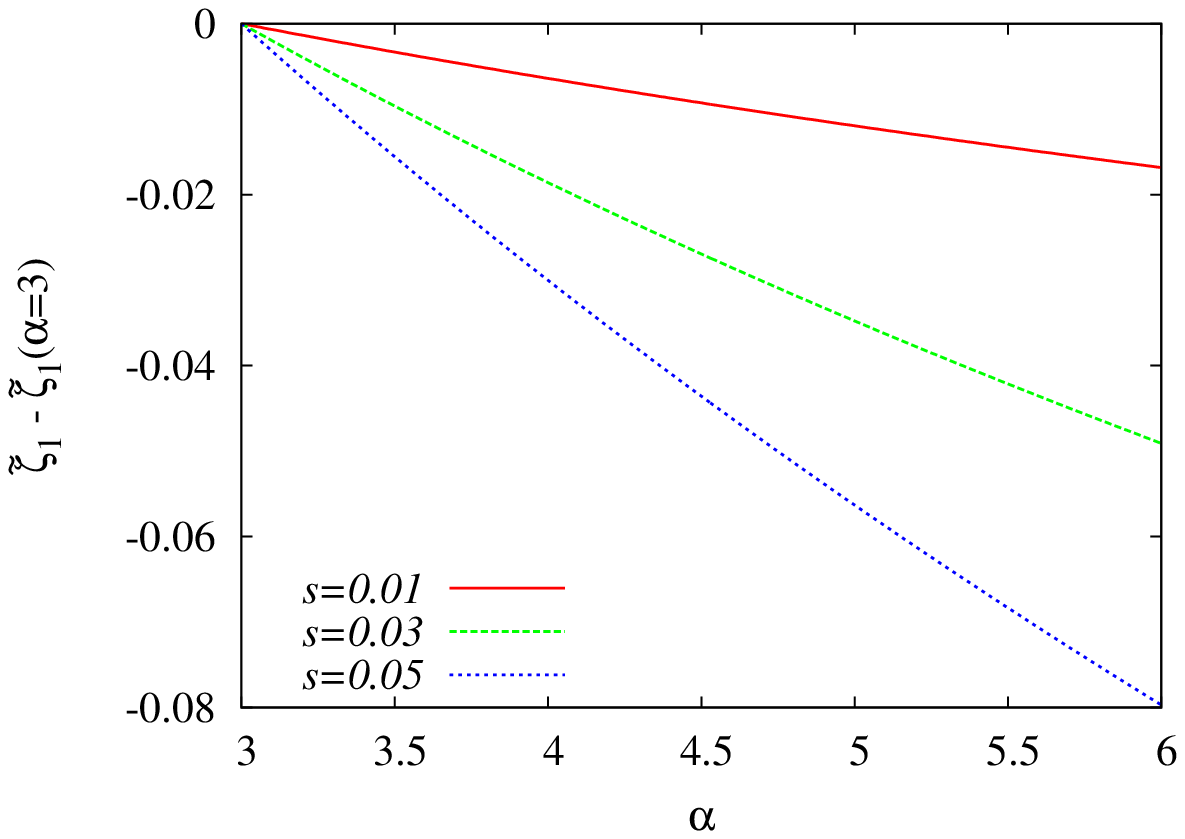}
    }
  \end{center}
  \caption{Plots of $\tilde{\zeta}_1 = \zeta_1 /\zeta_1^{\rm
      (quadratic)}$, which is normalized to $\zeta_1$ for the pure
    quadratic case, relative to that for the cases with $\alpha =4$
    (left) and $\alpha =3$ (right).  The values of $n$ and $r_{\rm
      dec}$ are taken as $n=8$ and $r_{\rm dec}=0.01$.}
  \label{fig:zeta_2}
\end{figure}

\begin{figure}[htbp]
  \begin{center}
    \resizebox{170mm}{!}{
        \hspace{-20mm}
    \includegraphics{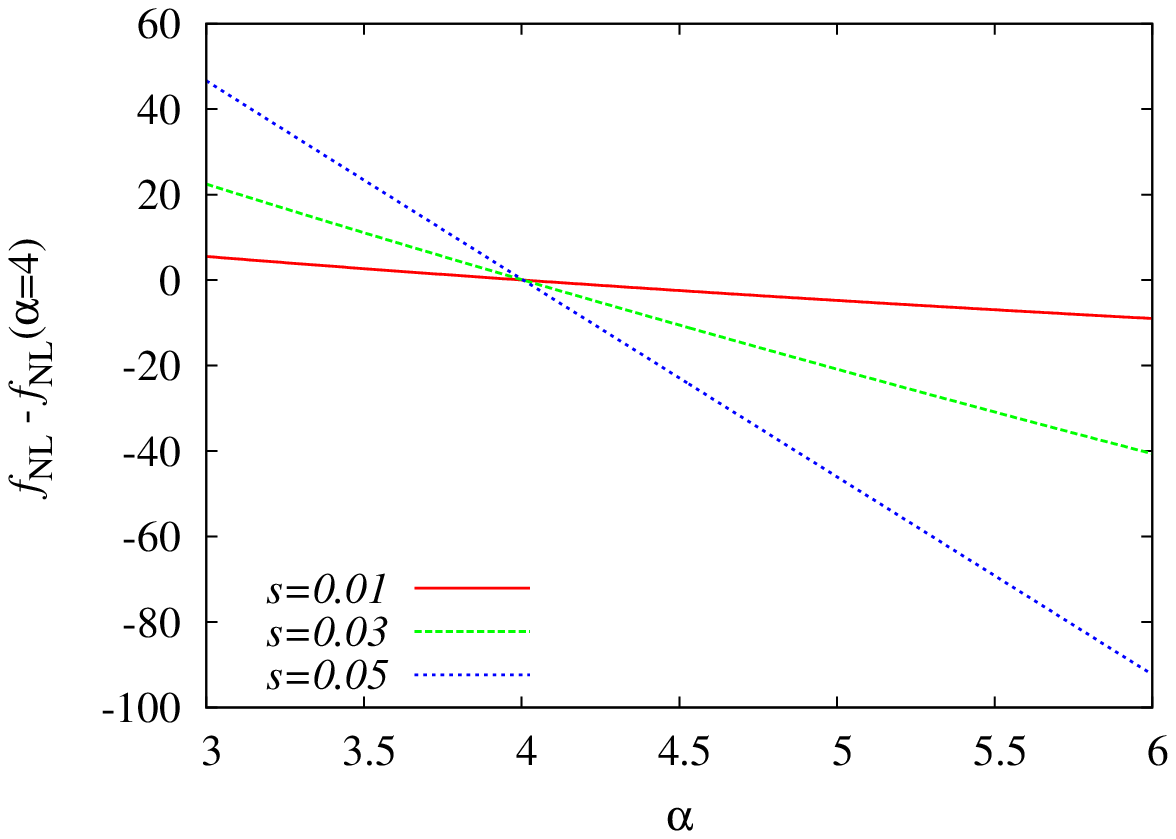}
    \includegraphics{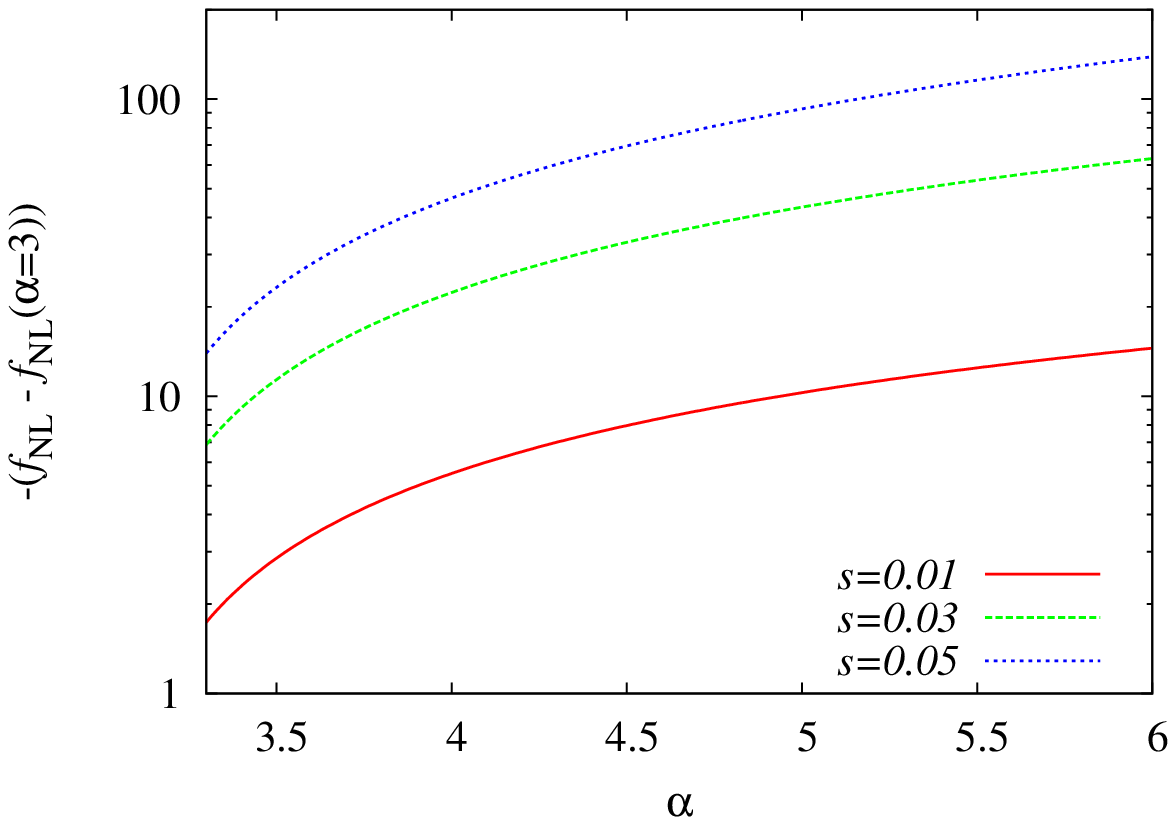}
    }
  \end{center}
  \caption{Plots of $f_{\rm NL}$ relative to that for the cases with
    $\alpha =4$ (left) and $\alpha =3$ (right).  The values of $n$ and
    $r_{\rm dec}$ are taken as $n=8$ and $r_{\rm dec}=0.01$.}
  \label{fig:fNL_2}
\end{figure}

\begin{figure}[htbp]
  \begin{center}
    \resizebox{170mm}{!}{
        \hspace{-20mm}
    \includegraphics{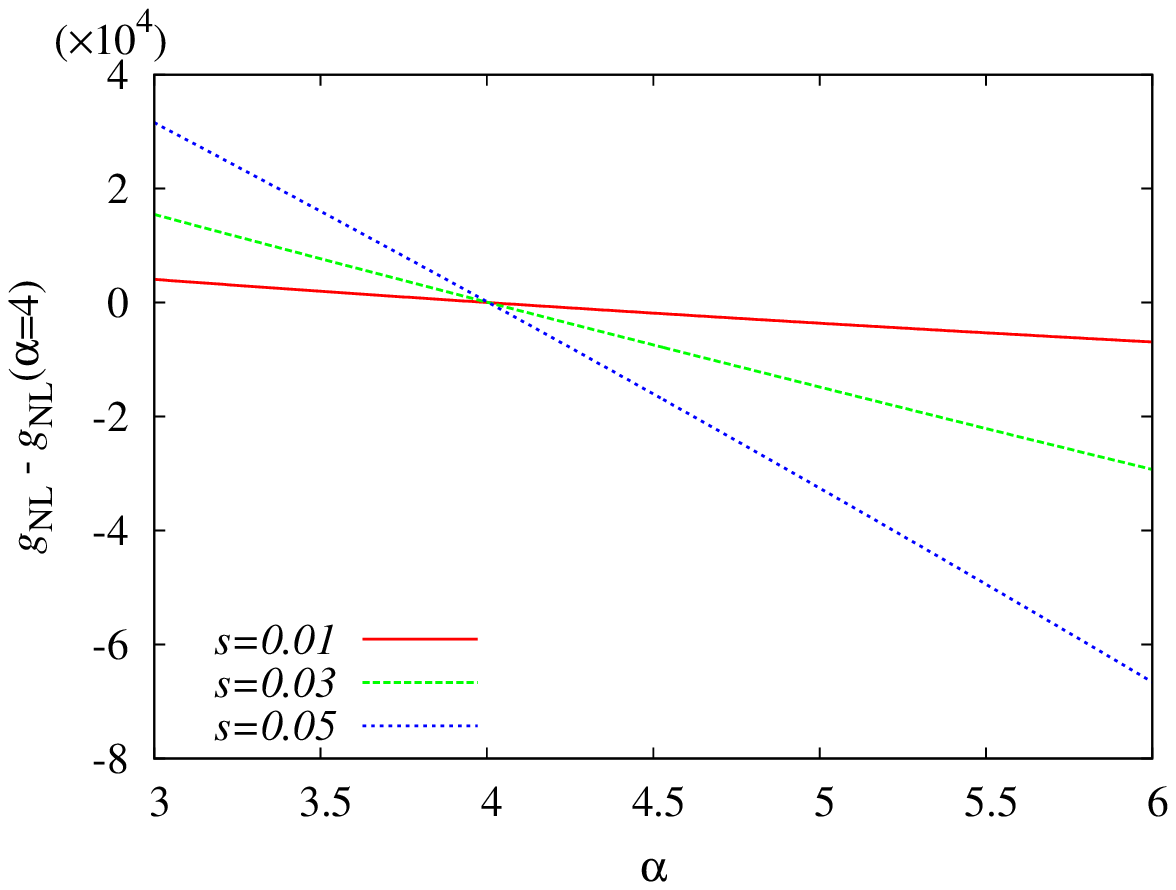}
    \includegraphics{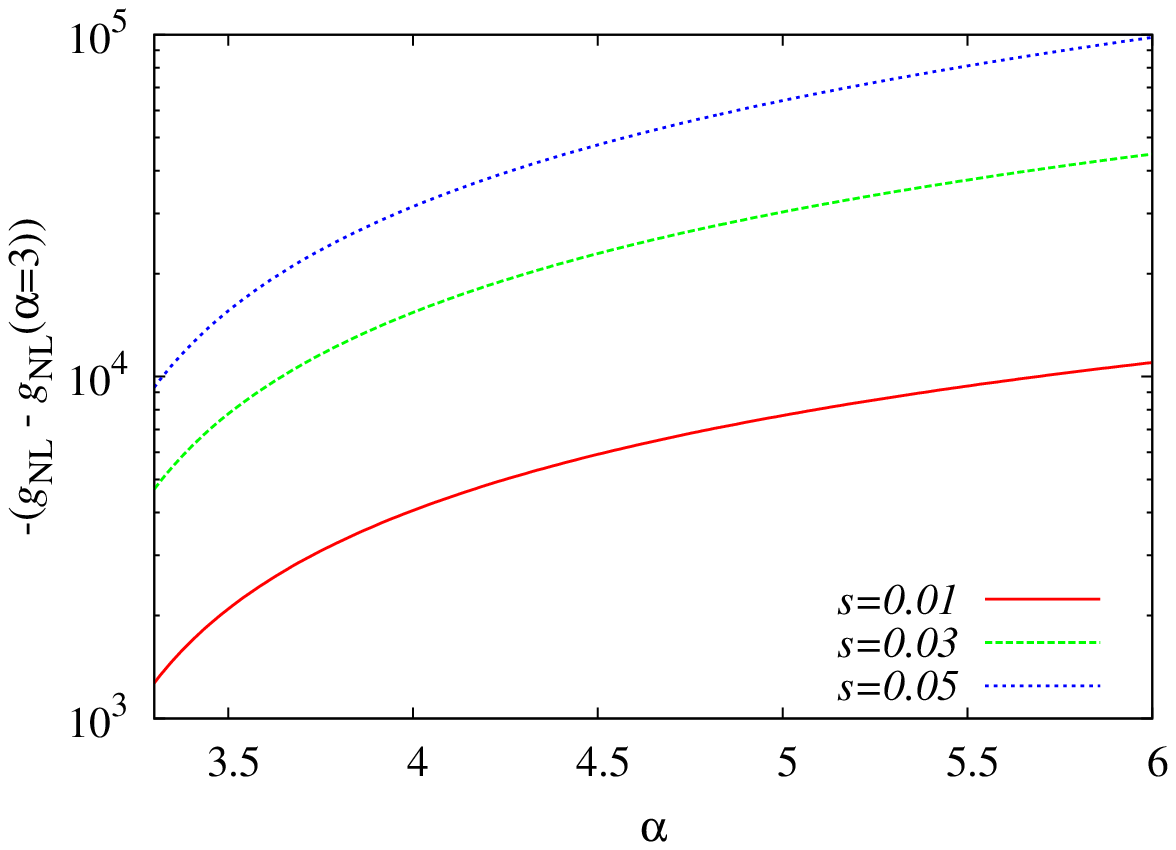}
    }
  \end{center}
  \caption{Plots of $g_{\rm NL}$ relative to that for the cases with
    $\alpha =4$ (left) and $\alpha =3$ (right).  The values of $n$ and
    $r_{\rm dec}$ are taken as $n=8$ and $r_{\rm dec}=0.01$.}
      \label{fig:gNL_2}
\end{figure}

\begin{figure}[htbp]
  \begin{center}
    \resizebox{100mm}{!}{
    \includegraphics{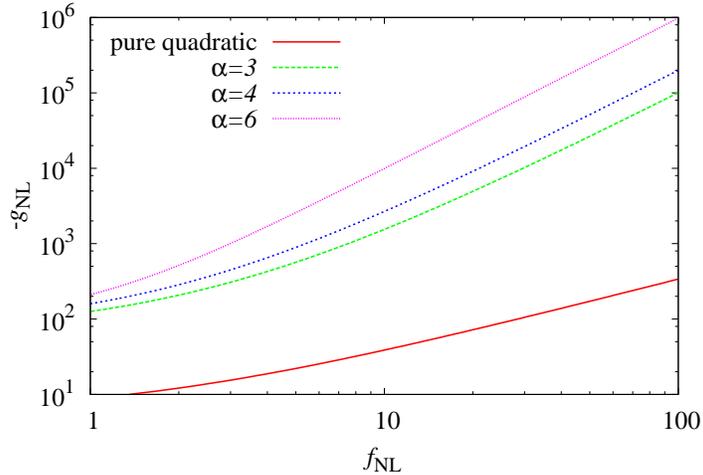}
    }
  \end{center}
  \caption{A relation between $f_{\rm NL}$ and $g_{\rm NL}$ for the
    case with $\alpha =3, 4$ and $6$.  The values of $s$ and $n$ are
    taken as $s=0.02$ and $n=8$. For reference, a pure quadratic case
    with $\alpha=4$ is also shown.}
  \label{fig:fNL_gNL}
\end{figure}

When there is no transition in the background evolution we can adopt
the formulae Eqs.~\eqref{eq:zeta_apprx}, \eqref{eq:fNL_apprx} and
\eqref{eq:gNL_apprx} by replacing $\beta$ by $\alpha$ in the
equations, as mentioned above.  To appreciate the impact of background
evolution on the curvature perturbation and its nonlinearity, we plot
$\zeta_1, f_{\rm NL}$ and $g_{\rm NL}$ in
Figs.~\ref{fig:fNL_gNL_n}--\ref{fig:fNL_gNL}. In
Fig.~\ref{fig:fNL_gNL_n}, the values of $f_{\rm NL}$ and $g_{\rm NL}$
are shown as a function of $n$ (see Eq. (\ref{eq:def_s}) for the
definition of the dimension of the non-quadratic contribution to the
potential) for the cases of $\alpha = 3, 4$ and $6$ which correspond
to matter, radiation and kination background, respectively, in our
notation.  As seen from the figures, the larger the non-quadratic
power $n$ is, the more pronounced the effect of the background
becomes.

To observe the magnitude of the effect more quantitatively, the values
of $\zeta_1, f_{\rm NL}$ and $g_{\rm NL}$ relative to those for the
case with $\alpha=4$ and $3$ are plotted in
Figs.~\ref{fig:zeta_1}--\ref{fig:gNL_1} for several values of $n$ as a
function of $\alpha$.  In these figures, we fix the relative strength
of the non-quadratic part $s$ and $r_{\rm dec}$ to $s=0.05$ and
$r_{\rm dec}=0.01$.  In Figs.~\ref{fig:zeta_2}--\ref{fig:gNL_2} we
also plot the same information but this time for several values of
$s$, fixing $n=8$ and $r_{\rm dec}$.

Interestingly, when the potential of the curvaton is close to a purely
quadratic one, the background evolution does not affect the results
much.  However, whenever the potential deviates from the quadratic
form, the background tends to suppress the curvature perturbation
$\zeta_1$ the more non-quadratic the potential is.  It should also be
noted that when $\alpha$ increases, or the background fluid becomes
more stiff, $\zeta_1$ becomes suppressed. For fixed $s$ and $n$, the
stiffening of the background fluid drives $f_{\rm NL}$ and $g_{\rm
  NL}$ to increasingly negative territory. A point worth stressing is
that these modifications are not small but e.g. by changing of the
radiation dominated background to matter dominated one induces shifts
in the non-linearity parameters that in principle could be easily
observable.

Finally, let us comment on the relation between $f_{\rm NL}$ and
$g_{\rm NL}$.  As pointed out in \cite{Enqvist:2008gk}, when the
potential of the curvaton has a purely quadratic form and $r_{\rm
  dec}$ is small, $f_{\rm NL}$ and $g_{\rm NL}$ are related as
\begin{equation}
g_{\rm NL} \simeq -\frac{10}{3} f_{\rm NL}.
\end{equation}
However, when the potential deviates from a quadratic form, ``the
consistency relation" between $f_{\rm NL}$ and $g_{\rm NL}$ becomes
\begin{equation}
\label{eq:nonQ_relation}
g_{\rm NL} \simeq \frac{3}{2} f_{\rm NL}^2
\left(
\frac{\sigma_{\rm osc}^2 \sigma_{\rm osc}^{\prime\prime\prime}}
{\sigma_{\rm osc}^{\prime 3}}
+
3\frac{\sigma_{\rm osc} \sigma_{\rm osc}^{\prime\prime}}{\sigma_{\rm osc}^{\prime 2}}
\right)
\left(
1 +
\frac{\sigma_{\rm osc} \sigma_{\rm osc}^{\prime\prime}}{\sigma_{\rm osc}^{\prime 2}}
\right)^{-1},
\end{equation}
where $r_{\rm dec} \ll 1$ is assumed. Because of the nonlinear
evolution of $\sigma$, which is encoded in $\sigma_{\rm osc}$ and its
derivatives appearing on the right hand side of
Eq.~\eqref{eq:nonQ_relation}, the nonlinearity parameters are now
related as $-g_{\rm NL} \propto f_{\rm NL}^2$. This is in sharp
contrast to the quadratic case where $-g_{\rm NL} \propto f_{\rm NL}$.
Furthermore, since the coefficient of the relation depends on the
nonlinear evolution of $\sigma$, it can be affected by the form of the
potential and the background evolution.  In this respect, the
comparison of $f_{\rm NL}$ and $-g_{\rm NL}$, if they ever are
observed\footnote{
  Recently, a constraint on $g_{\rm NL}$ has been obtained for the
  case of a negligible $f_{\rm NL}$; the limit is $-3.5 \times 10^5 <
  g_{\rm NL} < 8.2 \times 10^5$ \cite{Desjacques:2009jb}.
}, can be very useful for probing the form of the curvaton potential
and the equation of state of the background during the time when the
curvaton fluctuations generate the curvature perturbation.

To demonstrate this explicitly, in Fig.~\ref{fig:fNL_gNL} we plot the
value of $g_{\rm NL}$ as a function of $f_{\rm NL}$ for several values
of $\alpha$ with $n=8$ and $s=0.02$ . For reference, we also plot the
case of the pure quadratic potential with RD background
$(\alpha=4)$. When $f_{\rm NL}$ is large, which corresponds to the
case of $r_{\rm dec} \ll 1$, the above relations hold. As a
consequence, as one can see in Fig.~\ref{fig:fNL_gNL}, there is a
definite difference in the $g_{\rm NL}- f_{\rm NL}$ relation for
different backgrounds even when the non-quadratic term in the curvaton
potential remains the same. Thus the trispectrum may also provide
important information not only about the curvaton potential but also
about the equation of state of the background fluid during curvaton
oscillations.



\pagebreak

\section{Conclusion}\label{sec:conclusion}

In this paper, we have investigated the effect of the background
evolution on the curvaton non-Gaussianity, assuming a curvaton
potential which slightly deviates from the quadratic one. Such a study
is motivated by the possibility that after inflation, the inflaton
keeps oscillating about its global minimum for a long time so that the
curvaton could actually decay while the universe is still effectively
matter dominated. More exotic temporary possibilities, such as a
kination driven universe, could also be envisaged. Therefore we have
considered an ideal background fluid with some generic equation of
state, leading to a background evolution $\propto a^{-\alpha}$, with
$\alpha$ a free parameter.

It turns out that the changing of the background to radiation, or
equivalently, a change in the value of $\alpha$, that takes place
during curvaton oscillations, has by itself little effect on the
perturbation. What matters is the nature of the background evolution
before radiation domination finally kicks in, and as we show, it can
lead to significant and potentially observable consequences. The
non-linearity parameters $f_{\rm NL}$ and $g_{\rm NL}$, as well as the
linear curvature perturbation $\zeta_1$, depend on the nature of the
background fluid. We find that the dependence on the background fluid
becomes more pronounced as the deviation of the curvaton potential
from the quadratic one increases. Typically, when replacing one
background fluid with another, one induces effects on $f_{\rm NL}$
that are easily of the order of $\mathcal{O}(10)$ but could also be
much larger, depending on the relative strength of the non-quadratic
part of the potential.

An interesting issue is the relation between $f_{\rm NL}$ and $g_{\rm
  NL}$. We have showed that the relation depends both on the form of
the curvaton potential and the background evolution, which we find a
rather surprising result.  Hence measuring both $f_{\rm NL}$ and
$g_{\rm NL}$, or both the bispectrum and the trispectrum, would yield
information not only on the curvaton self-interactions but also on the
equation of state of the background fluid. Since that is linked to
dynamics in the inflaton sector, here arises a possibility of probing
inflaton physics at the very end of inflation.

\bigskip
\bigskip

\noindent {\bf Acknowledgments:} T.T. would like to thank the Helsinki
Institute of Physics for the hospitality during the visit, where this
work was initiated.  This work is supported in part the Grant-in-Aid
for Scientific Research from the Ministry of Education, Science,
Sports, and Culture of Japan No.\,19740145 (T.T.), and in part by the
Academy of Finland grant 114419 (K.E.).

\end{document}